\documentclass[11pt]{article}
\usepackage{moriond,epsfig}

\def\Journal#1#2#3#4{{#1} {\bf #2}, #3 (#4)}

\def\NPB{{\em Nucl. Phys.} B}

\def\PRD{{\em Phys. Rev.} D}

\def\JPG{{\em J. Phys.}~{G}}
\def\ARNPS{{\em Ann. Rev. Nucl. Part. Sci.\ }}
\def\NPA{{\em Nucl. Phys.} A}
\def\PRC{{\em Phys. Rev.} C}

\def\be{\begin{equation}}
\def\ee{\end{equation}}
\def\bea{\begin{eqnarray}}
\def\eea{\end{eqnarray}}

\def\mean#1{\left<#1\right>}
\begin{document}
\vspace*{4cm}
\title{WHERE FEYNMAN, FIELD AND FOX FAILED AND HOW WE FIXED IT AT RHIC}

\author{ M.J. TANNENBAUM }

\address{Physics Department, 510c, Brookhaven National Laboratory\\
Upton, NY 11973-5000, U.S.A.}

\maketitle\abstracts{
  Hard-scattering of point-like constituents (or partons) in p-p collisions was discovered at the CERN-ISR in 1972 by measurements utilizing inclusive single or pairs of hadrons with large transverse momentum ($p_T$). 
 It was generally assumed following a seminal paper by Feynman, Field and Fox (FFF)~\cite{FFF} (and much discussed in a talk that I gave at the 1979 Rencontres de Moriond) that ``everything you wanted to know about hard-scattering and jets'' could be measured by these methods. Recently, we found in PHENIX~\cite{ppg029} that the $p_{T_a}$ distribution of away side hadrons from a single particle trigger [with $p_{T_t}$] which is a leading fragment of the trigger jet, could not be used to  measure the fragmentation function of the away jet as originally claimed by FFF. A new formula was derived which both exhibits scaling in the variable $x_E\sim p_{T_a}/p_{T_t}$ (a  hot topic in 1979) and relates $x_E$, $\sim$ the ratio of the transverse momenta of the measured particles, to $\hat{x}_h=\hat{p}_{T_a}/\hat{p}_{T_t}$, the ratio of the transverse momenta of the away-side to trigger-side jets. Tests of the validity of the formula and applications to Au+Au central collisions at RHIC (where jets can not be reconstructed) are discussed. }    
 \section{Introduction}
\begin{figure}[!htb]%%\vspace*{-0.25cm}%\vspace*{-0.5cm}
\begin{center}
\begin{tabular}{ccc}
%\hspace*{-4cm}
\psfig{file=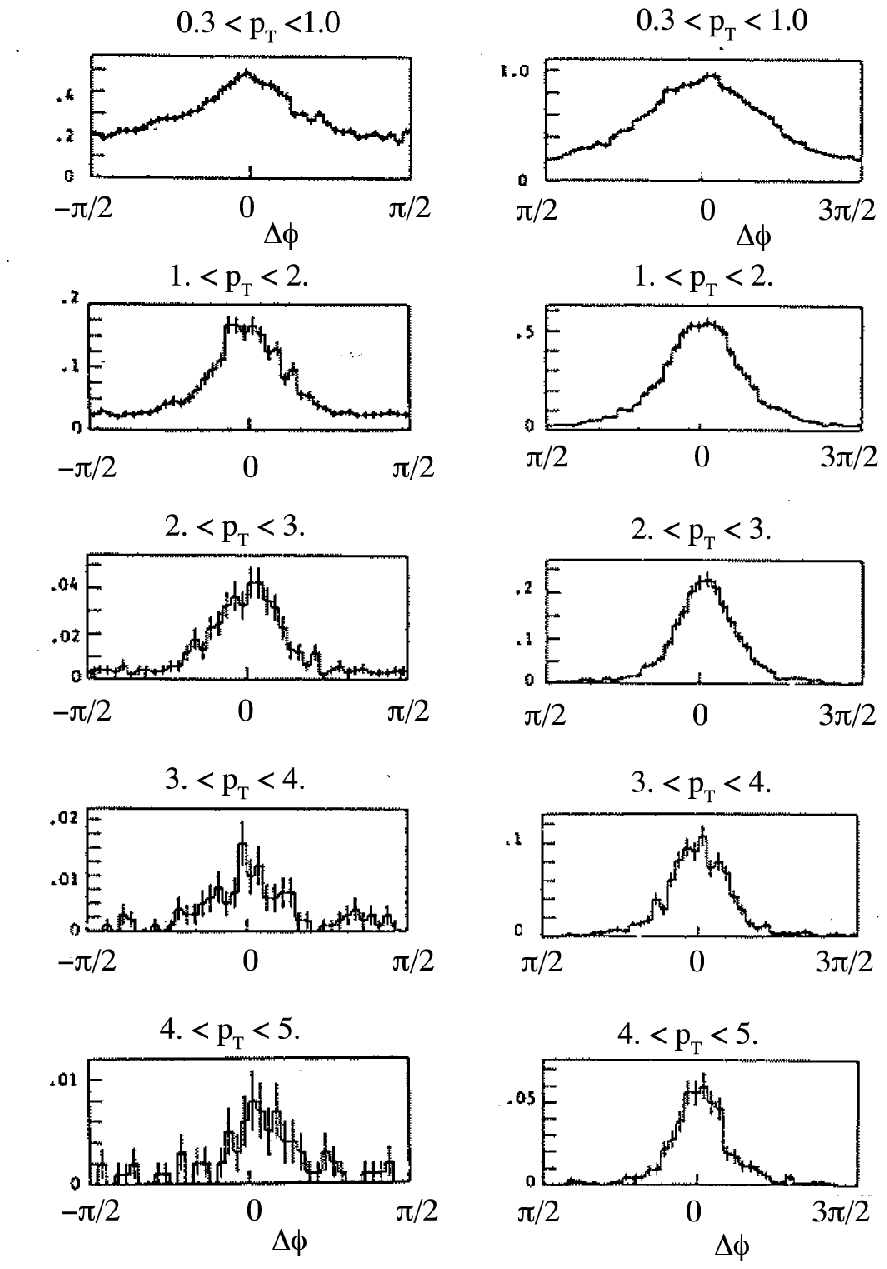,width=0.30\linewidth,height=0.36\linewidth}&%\hspace*{-0.044\linewidth}
\psfig{file=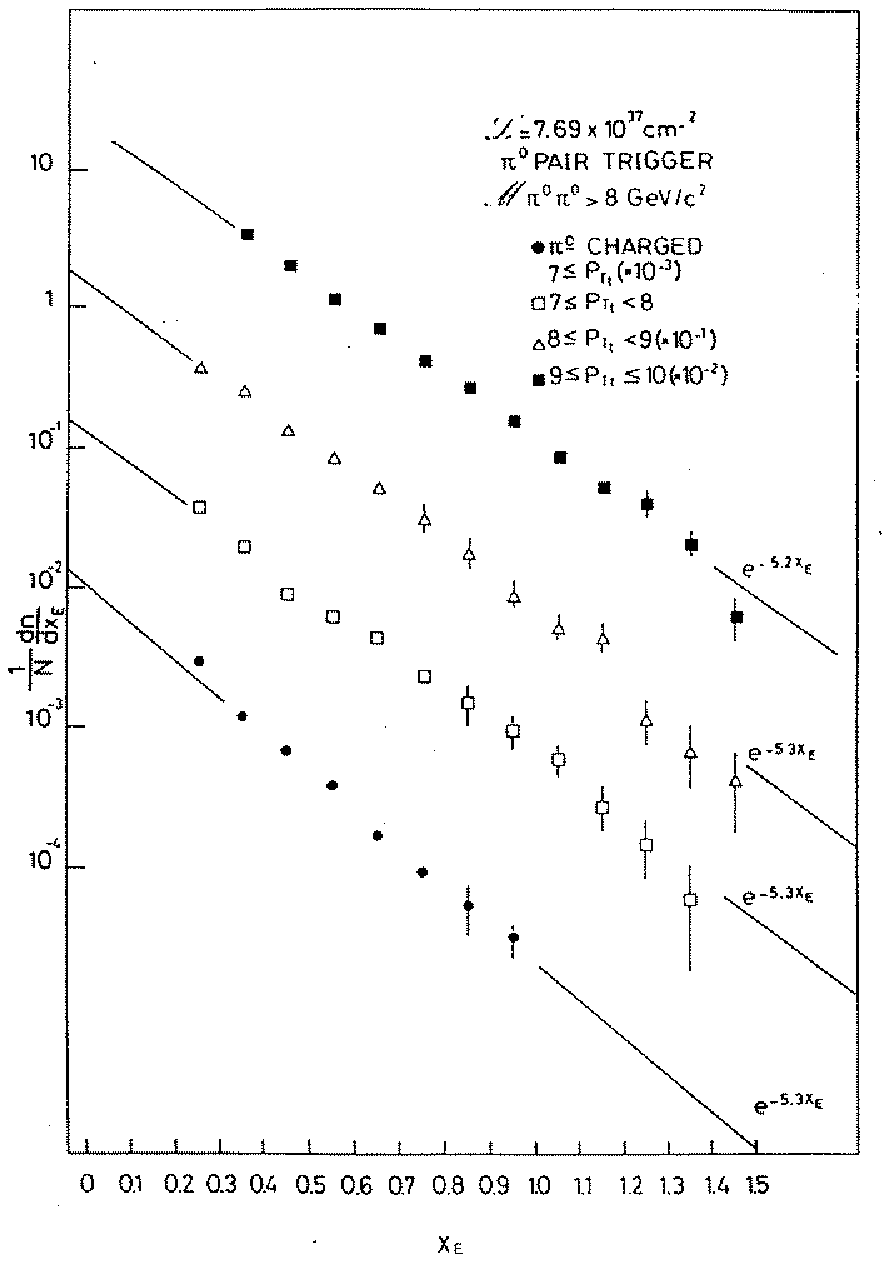,width=0.30\linewidth,height=0.36\linewidth}&
\psfig{file=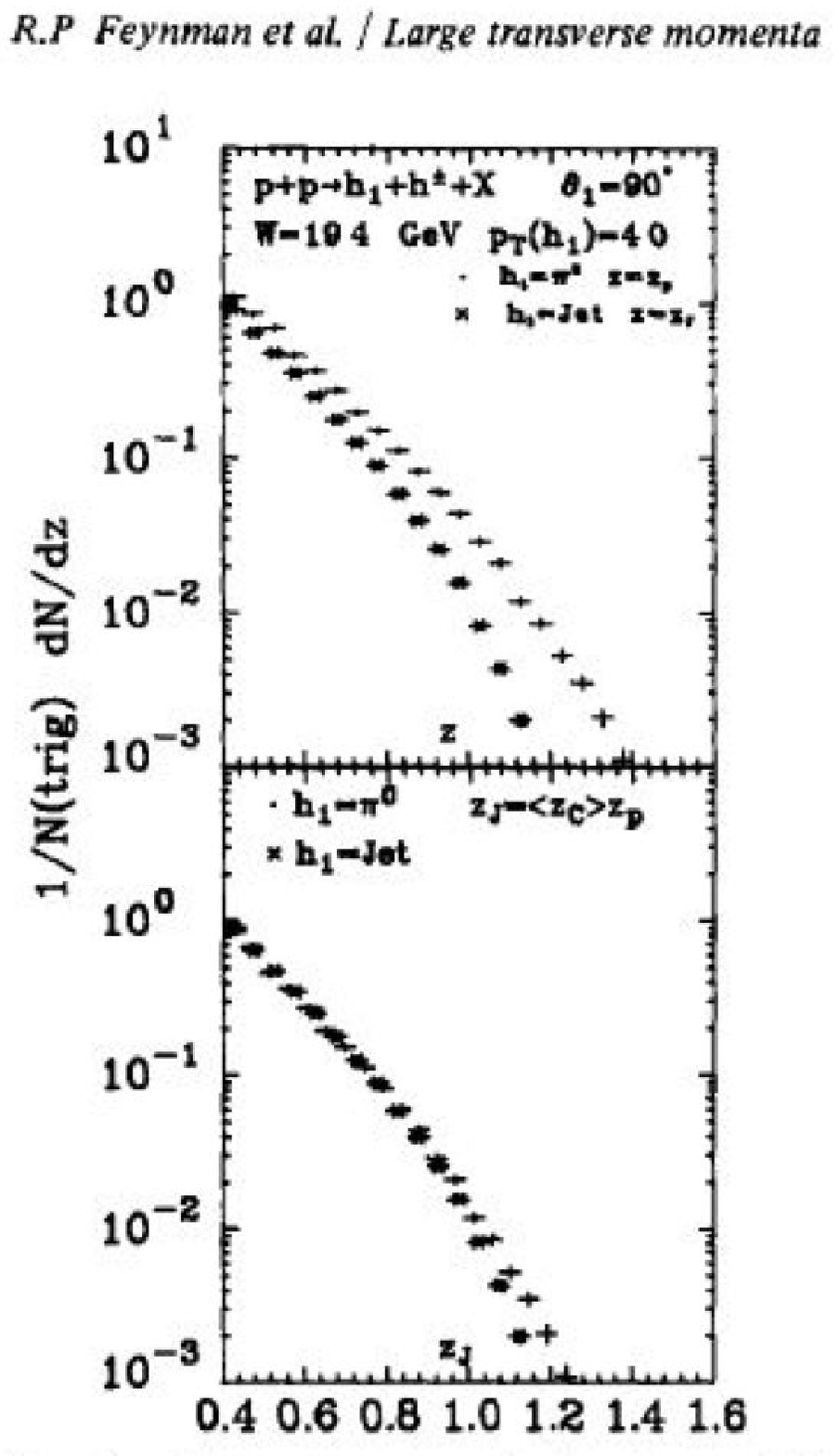,width=0.30\linewidth,height=0.36\linewidth}%%
\end{tabular}
\end{center}\vspace*{-0.15in}
\caption[]{a) (left) trigger-side b) (left-center) away-side correlations of charged particles with indicated $p_T\equiv p_{T_a}$ for $\pi^0$ triggers with $p_{T_t} > 7$ GeV/c  . c) (right-center) $x_E$ distributions from this data d) right) [top] Comparison~\cite{FFF} of away side charged hadron distribution triggered by a $\pi^0$ or a jet, where $z_{\pi^0}=x_E$ and $z_j=p_{T_a}/\hat{p}_{T_a}$. [bottom] same distributions with $\pi^0$ plotted vs $z'_j=\mean{z_t} x_E$.    \label{fig:mjt-ccorazi}}\vspace*{-0.15in}
\end{figure}

Following the discovery of hard-scattering in p-p collisions at the CERN-ISR~\cite{Darriulat} by the observation of an unexpectedly large yield of particles with large transverse momentum $(p_T)$, which proved that the quarks of DIS were strongly interacting, the attention of experimenters turned to measuring the predicted di-jet structure of the hard-scattering events. The jet structure of high $p_T$ scattering could be easily seen and measured using two-particle correlations, e.g. with a $\pi^0$ trigger with transverse momentum $p_{T_t} > 7$ GeV/c, and an associated charged particle detected with full and uniform acceptance over the entire azimuth, with pseudorapidity coverage $-0.7\leq\eta\leq +0.7$ (Fig.~\ref{fig:mjt-ccorazi}a,b)~\cite{Angelis79}. In all cases strong correlation peaks on flat backgrounds are clearly visible for both the trigger-side and the away-side, indicating di-jet structure.      
      	Following the methods of previous CERN-ISR experiments~\cite{Darriulat}  and the best theoretical guidance~\cite{FFF}, the away jet azimuthal angular distributions  of Fig.~\ref{fig:mjt-ccorazi}b, which were thought to be unbiased, were analyzed in terms of the two variables: $p_{\rm out}=p_{T_a} \sin(\Delta\phi)$, the out-of-plane transverse momentum of a track,   
 and $x_E$, where \\ 
\hspace*{0.05\linewidth}\begin{minipage}[b]{0.45\linewidth}
%\vspace*{-0.30in}
\[ %\begin{equation*}	
x_E=\frac{-\vec{p}_{T_a}\cdot \vec{p}_{T_t}}{|p_{T_t}|^2}=\frac{-p_{T_a} \cos(\Delta\phi)}{p_{T_t}}\simeq \frac {z_a}{z_{t}}  
%\label{eq:mjt-xE} 
%%\nonumber
%%\end{equation*}
\]
\vspace*{0.001in}
\end{minipage}
\hspace*{0.02\linewidth}
\begin{minipage}[b]{0.39\linewidth} 
%\begin{figure}[!h] %%was 0.55
\vspace*{0.003in}
\psfig{file=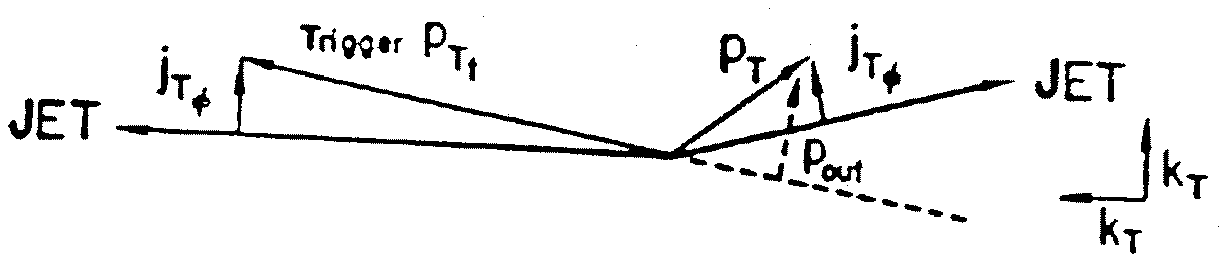, width=\linewidth}
\vspace*{-0.10in}
\label{fig:mjt-poutxe}
%\end{figure}
%\addtocounter{figure}{-1}
\end{minipage}%\vspace*{-0.12in}
\vspace*{-0.12in}

\noindent $z_{t}\simeq p_{T_t}/\hat{p}_{T_t}$ is the fragmentation variable of the trigger jet with $\hat{p}_{T_t}$, and $z_{a}\simeq p_{T_a}/\hat{p}_{T_a}$ is the fragmentation variable of the away jet. Note that $x_E$ would equal the fragmenation fraction $z_a$ of the away jet, for $z_{t}\rightarrow 1$, if the trigger and away jets balanced transverse momentum, i.e. if $\hat{x}_h\equiv\hat{p}_{T_a}/\hat{p}_{T_t}=1$.  
It was generally assumed, following the seminal article of Feynman, Field and Fox~\cite{FFF}, that the $p_{T_a}$ distribution of away side hadrons from a single particle trigger [with $p_{T_t}$], corrected for $\mean{z_t}$, would be the same as that from a jet-trigger (Fig.~\ref{fig:mjt-ccorazi}d) and follow the same fragmentation function as observed in $e^+ e^-$  or DIS~\cite{Darriulat}. The $x_E$ distributions~\cite{Angelis79} for the data of Fig.~\ref{fig:mjt-ccorazi}b are shown in Fig.~\ref{fig:mjt-ccorazi}c and show the fragmentation behavior expected at the time, $e^{-6\,z_{a}}\sim e^{-6\,\langle z_{t}\rangle\,  x_E }=e^{-5.3\,  x_E }$.\\[-0.35in] 

\section{Discoveries at RHIC}
\begin{figure}[!thb]
\begin{center}
\begin{tabular}{cc}
\includegraphics[width=0.38\linewidth]{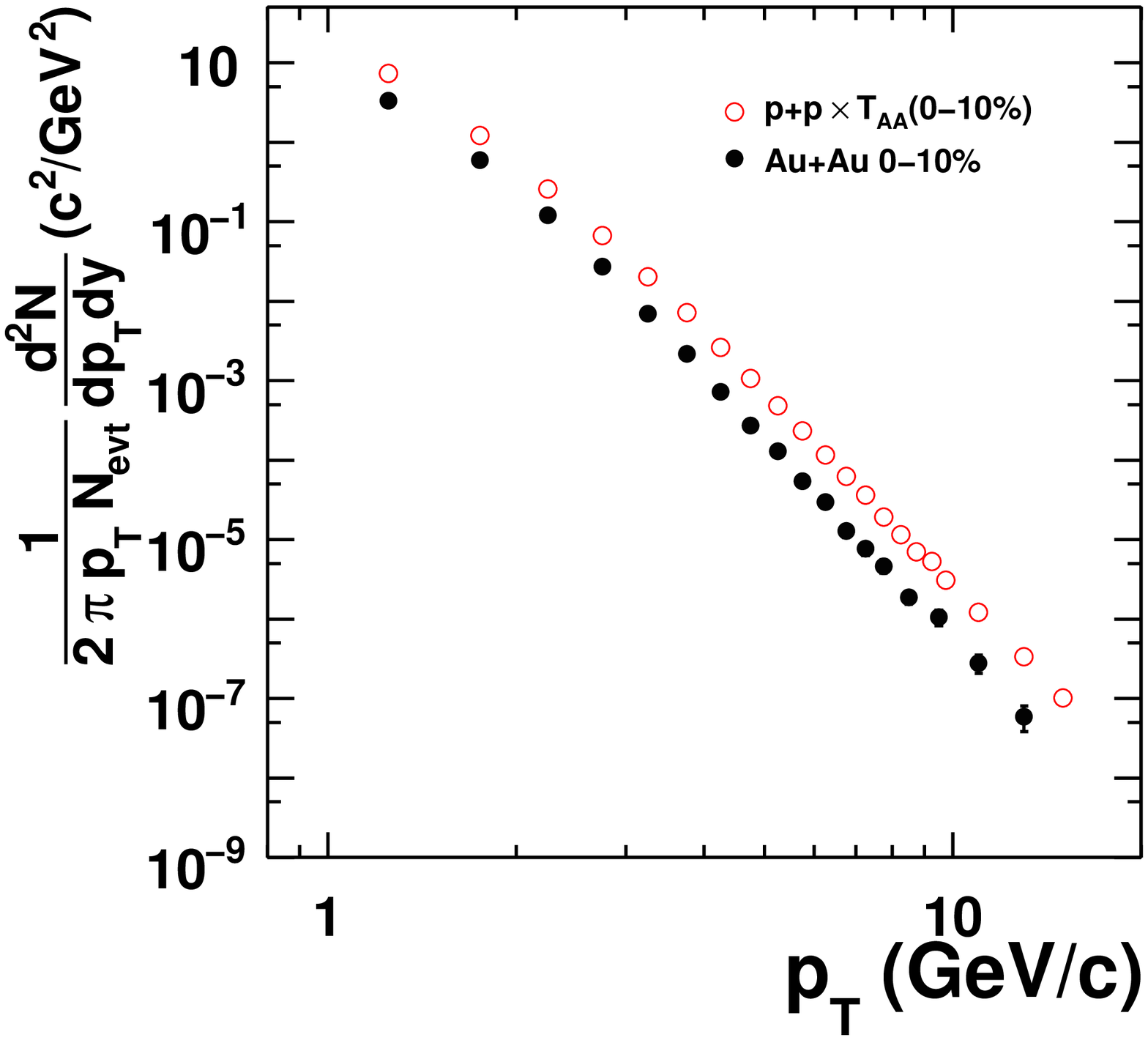}&
\hspace*{-0.02\linewidth}\includegraphics[width=0.50\linewidth]{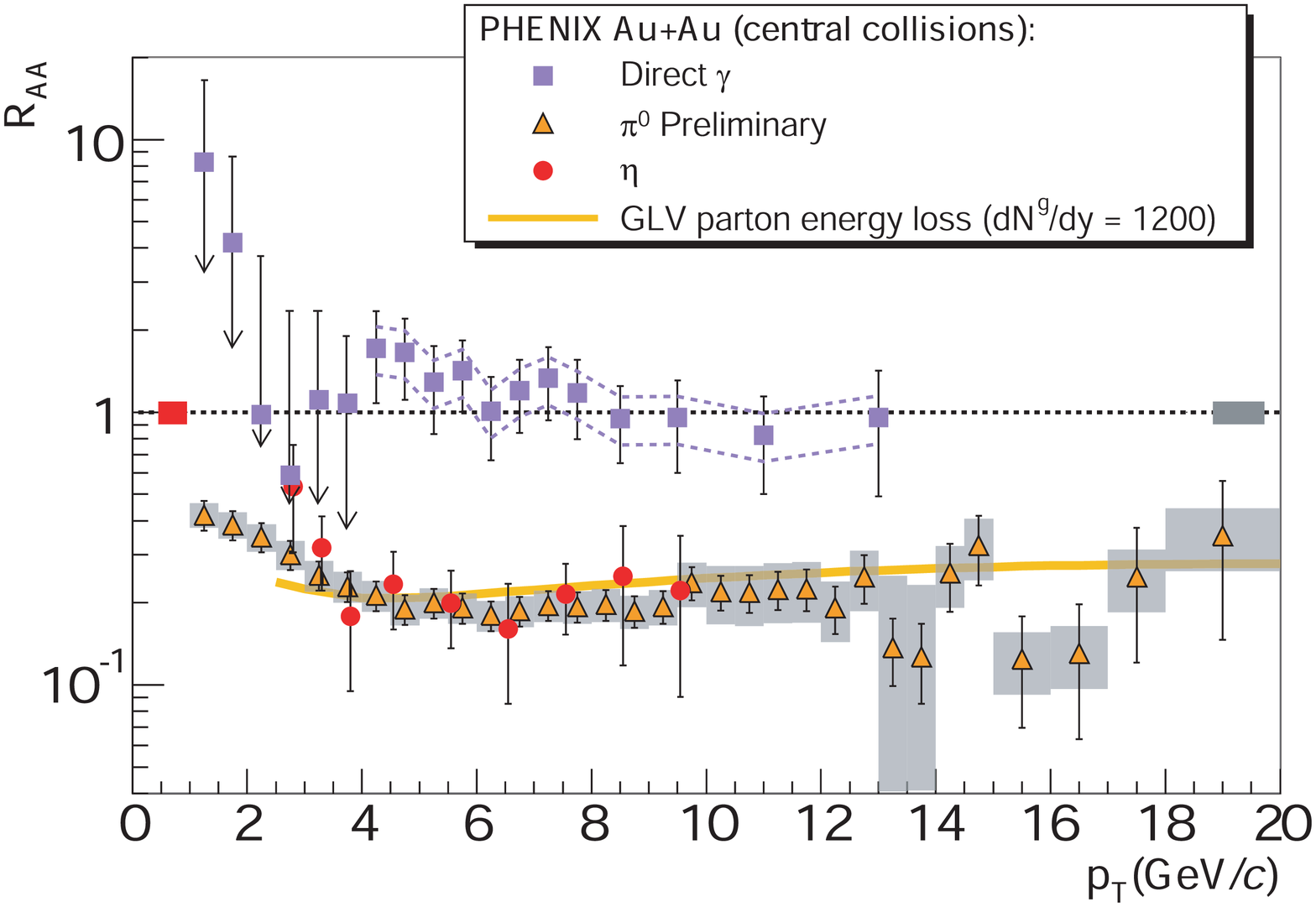}
\end{tabular}
\end{center}\vspace*{-0.25in}
\caption[]{(left)log-log plot~\cite{ppg054} of $\pi^0$ invariant cross section in p-p collisions at $\sqrt{s}=200$ GeV multiplied by $\mean{T_{AA}}$ for Au+Au central collisions (0-10\%) compared to the measured semi-inclusive invariant yield of $\pi^0$. (right) $R_{AA}(p_T)$ for $\pi^0$, $\eta$ and direct-$\gamma$  for Au+Au central (0-10\%) collisions at $\sqrt{s_{NN}}=200$ GeV~\cite{Wolf}.  }
\label{fig:RAA}\vspace*{-0.12in}
\end{figure}

        It was discovered at RHIC~\cite{egPXWP} that the high $p_T$ $\pi^0$ from hard-scattering are suppressed by roughly a factor of 5 in central Au+Au collisions compared to the scaling for point-like processes. This is arguably {\em the}  major discovery in Relativistic Heavy Ion Physics. The suppression is attributed to energy-loss of the outgoing hard-scattered color-charged partons due to interactions in the presumably deconfined and thus color-charged medium produced in Au+Au collisions at RHIC~\cite{BSZARNPS}. In Fig.~\ref{fig:RAA}-(left), a log-log plot of the $\pi^0$ invariant cross section in p-p collisions at $\sqrt{s}=200$ GeV multiplied by the point-like scaling factor $\mean{T_{AA}}$ (the overlap integral of the nuclear thickness functions averaged over the centrality class) for Au+Au central collisions (0-10\%) is compared to the measured semi-inclusive invariant yield of $\pi^0$. Both the Au+Au and p-p data show a pure power law, $p_T^{-8.10}$ for $p_T > 3$ GeV/c. The suppression is shown more dramatically in Fig.~\ref{fig:RAA}-(right) where the the data for $\pi^0$, $\eta$ and direct-$\gamma$ are presented as the ratio of the yield per central Au+Au collision  (upper 10\%-ile of observed multiplicity) to the point-like-scaled p-p cross section:
$  R_{AA}(p_T)={({d^2N^{\pi}_{AA}/N_{AA} dp_T dy})/ ({\langle T_{AA}\rangle d^2\sigma^{\pi}_{pp}/dp_T dy}})$ . 
The $\pi^0$ and $\eta$, which are fragments of hard-scattered partons, are both suppressed by a factor of 5 from $3\leq p_T\leq 20$ GeV/c. This implies a strong medium effect in Au+Au central collisions because the direct-$\gamma$, which also participate directly in the 2-to-2 hard-scattering but do not interact with the medium, are not suppressed. 
 
\begin{figure}[!htb]
\vspace*{-0.12in}\begin{center}
\begin{tabular}{ccc}
\psfig{file=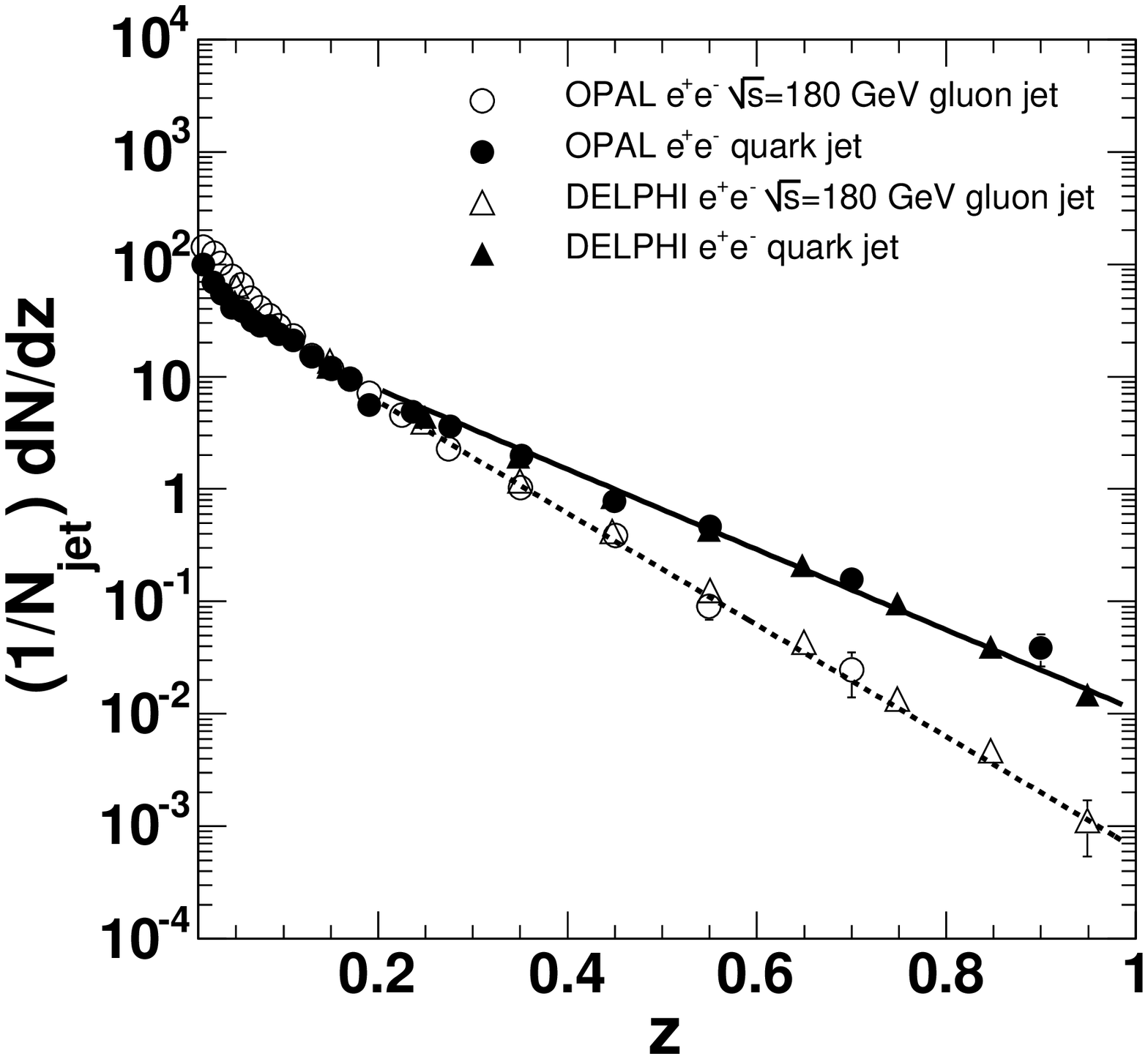,width=0.33\linewidth}&\hspace*{-0.053\linewidth}
\psfig{file=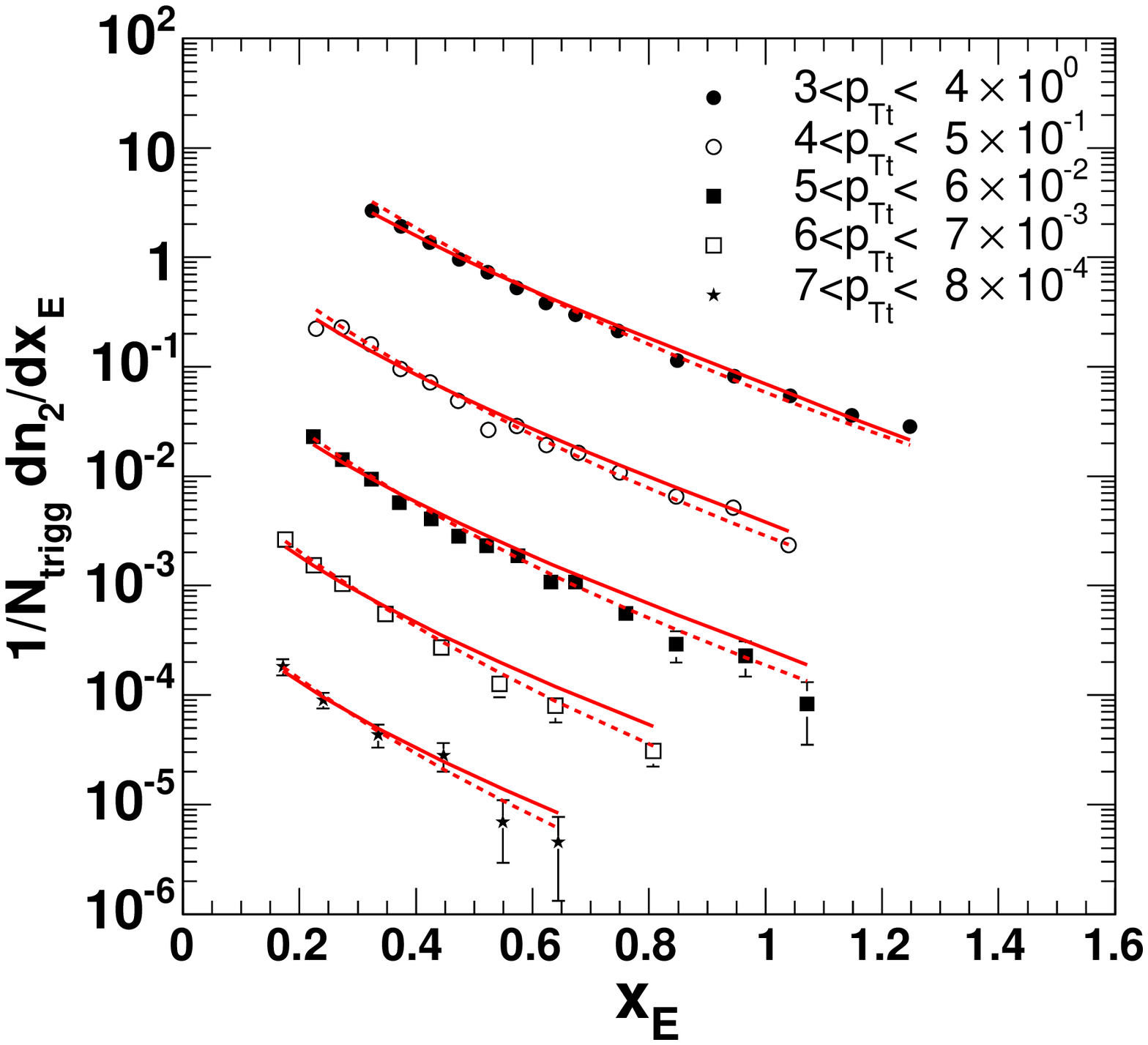,width=0.36\linewidth}&\hspace*{-0.06\linewidth}
\psfig{file=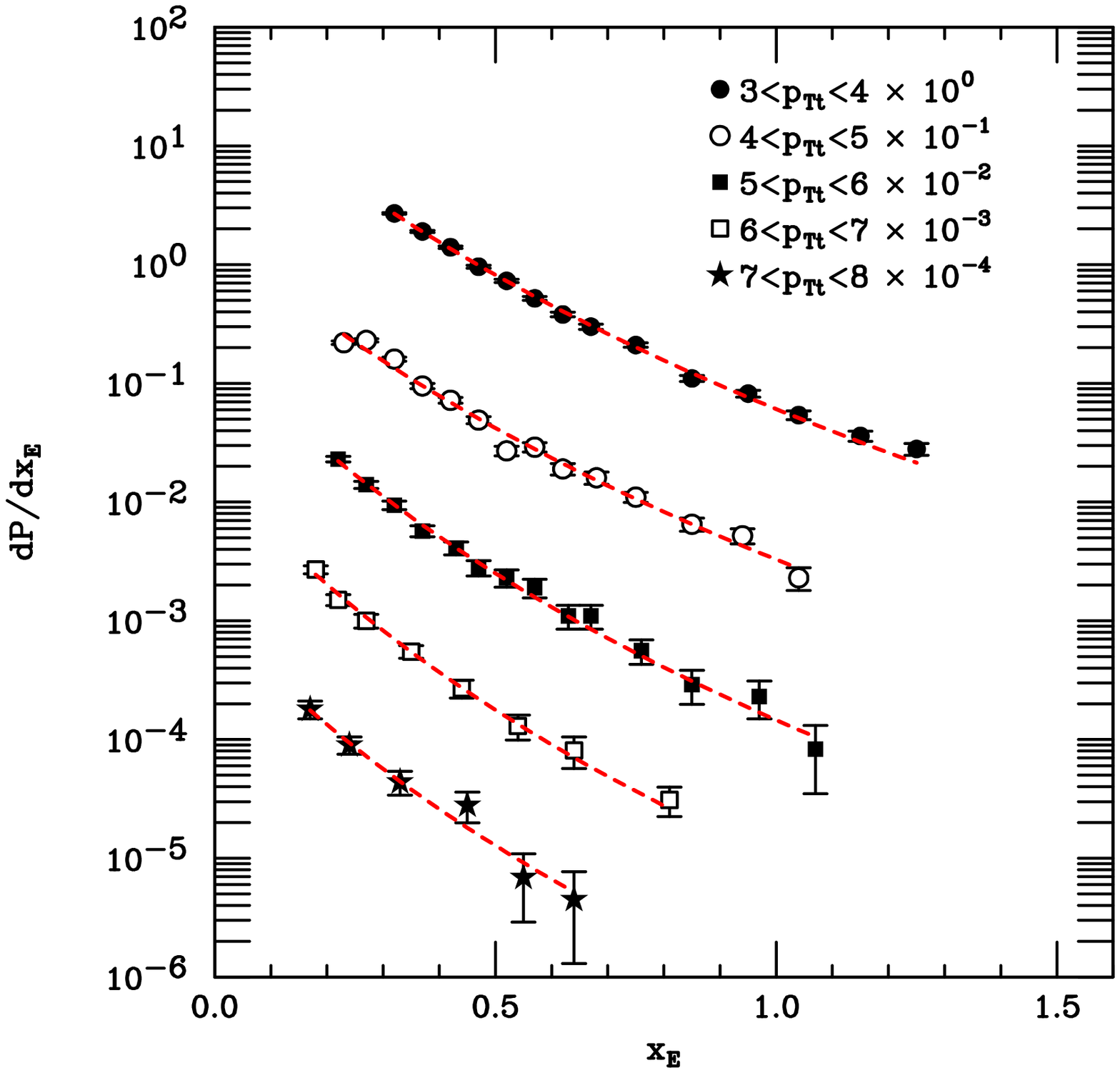,width=0.32\linewidth}
\end{tabular}
\end{center}\vspace*{-0.25in}
\caption[]{a) (left) LEP fragmentation functions. See Ref.~\cite{ppg029} for details; b) (center) $x_E$ distributions~\cite{ppg029} together with calculations using fragmentation functions from LEP; c) (right) data from (b) with fits to Eq.~\ref{eq:condxe2} for $n=8.10$.    \label{fig:xxx3}}\vspace*{-0.02in}
\end{figure}

    In order to measure whether the away-side parton from a high $p_T$ $\pi^0$ trigger loses energy in the medium formed in Au+Au central collisions or has modified fragmentation, 
    PHENIX~\cite{ppg029} attempted to derive the the fragmentation function in p-p collisions from the measured $x_E$ distributions (Fig.~\ref{fig:xxx3}b) to use as a baseline for the Au+Au measurement. 
It didn't work. Finally, it was found that starting with either the quark $\approx \exp (-8.2 \cdot z)$ or the gluon $\approx \exp (-11.4 \cdot z)$ fragmentation functions from LEP (Fig.~\ref{fig:xxx3}a solid and dotted lines), which are quite different in shape, the results obtained for the $x_E$ distributions (solid and dotted lines on Fig.~\ref{fig:xxx3}b) do not differ significantly! Although nobody had noticed this for nearly 30 years, the reason turned out to be quite simple~\cite{ppg029}. With no assumptions other than a power law for the jet $\hat{p}_{T_t}$ distribution (${{d\sigma_{q} }/{\hat{p}_{T_t} d\hat{p}_{T_t}}}= A \hat{p}_{T_t}^{-n}$), an exponential fragmentation function ($D^{\pi}_q (z)=B e^{-bz}$), and constant $\hat{x}_h$, for fixed $p_{T_t}$ as a function of $p_{T_a}$, it was possible~\cite{ppg029} to derive the $x_E$ distribution in the collinear limit, where $p_{T_a}=x_E p_{T_t}$: 
	     \begin{equation}
\left.{dP_{\pi} \over dx_E}\right|_{p_{T_t}}\approx {N (n-1)}{1\over\hat{x}_h} {1\over
{(1+ {x_E \over{\hat{x}_h}})^{n}}} \, \qquad ,  
\label{eq:condxe2}
\end{equation}
and $N=\mean{m}$ is the multiplicity of the unbiased away-jet. The shape of the $x_E$ distribution is given by the power $n$ of the partonic and inclusive single particle transverse momentum spectra and does not depend on the exponential slope of the fragmentation function as shown by the excellent fits of Eq.~\ref{eq:condxe2} (Fig.~\ref{fig:xxx3}c). Note that Eq.~\ref{eq:condxe2} provides a relationship between the ratio of the transverse momenta of the away to the trigger particles, $x_{E}\approx p_{T_a}/p_{T_t}$, which is measured, to the ratio of the transverse momenta of the away to the trigger jets, $\hat{x}_h=\hat{p}_{T_a}/\hat{p}_{T_t}$, which can thus be deduced. In p-p collisions, the imbalance of the away-jet and the trigger jet indicated by fitted values of ($\hat{x}_h\sim 0.7-0.9$) in Fig.~\ref{fig:xxx3}c is caused by $k_T$-smearing (the main topic of my 1979 Moriond talk).  In A+A collisions, $\hat{x}_h$ is sensitive to the relative energy loss of the trigger and associated jets in the medium, which can be thus measured.  
   \begin{figure}[!t]
\begin{center}
\begin{tabular}{ccc}
\includegraphics[width=0.36\linewidth]{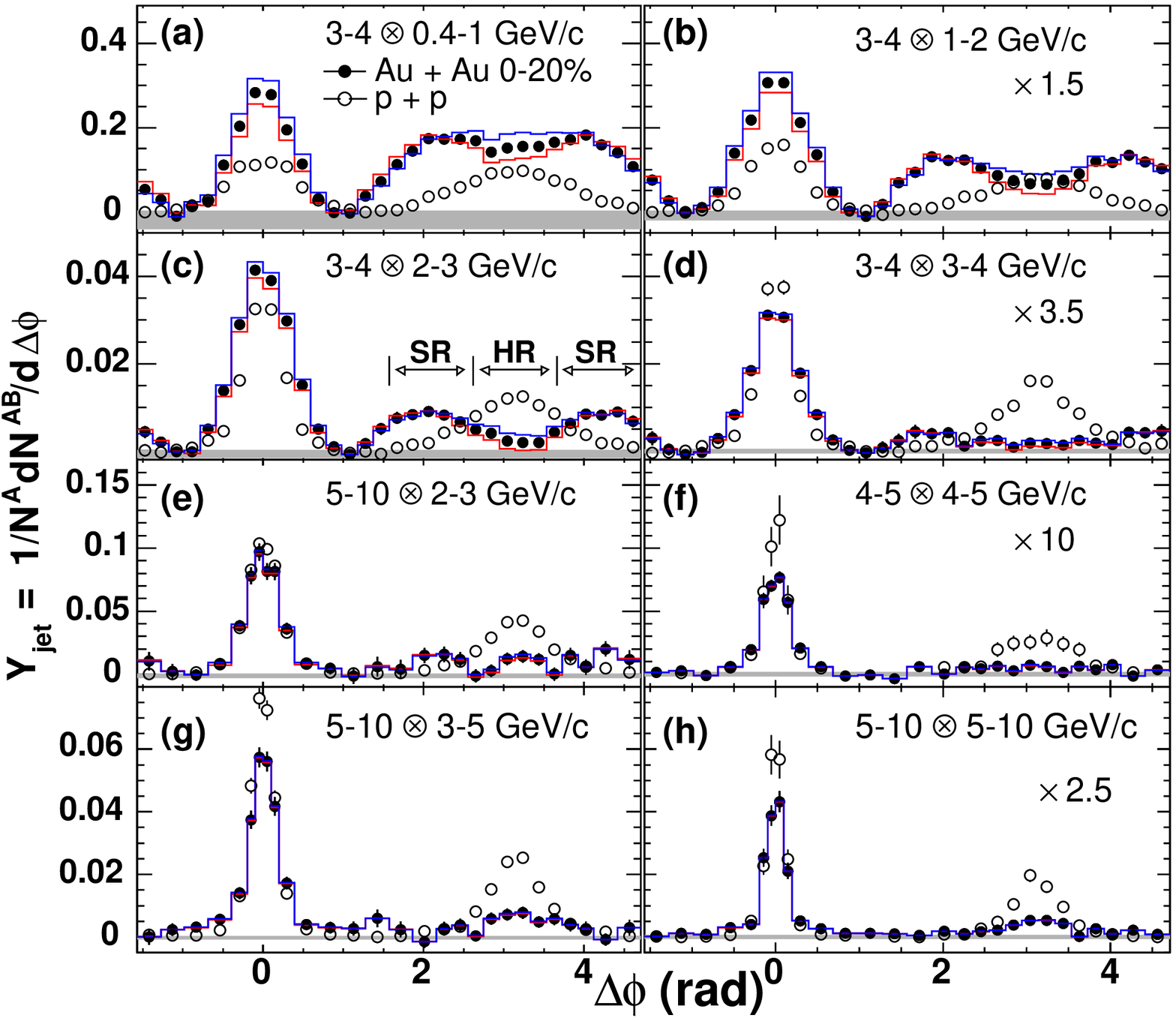}&
\includegraphics[width=0.28\linewidth]{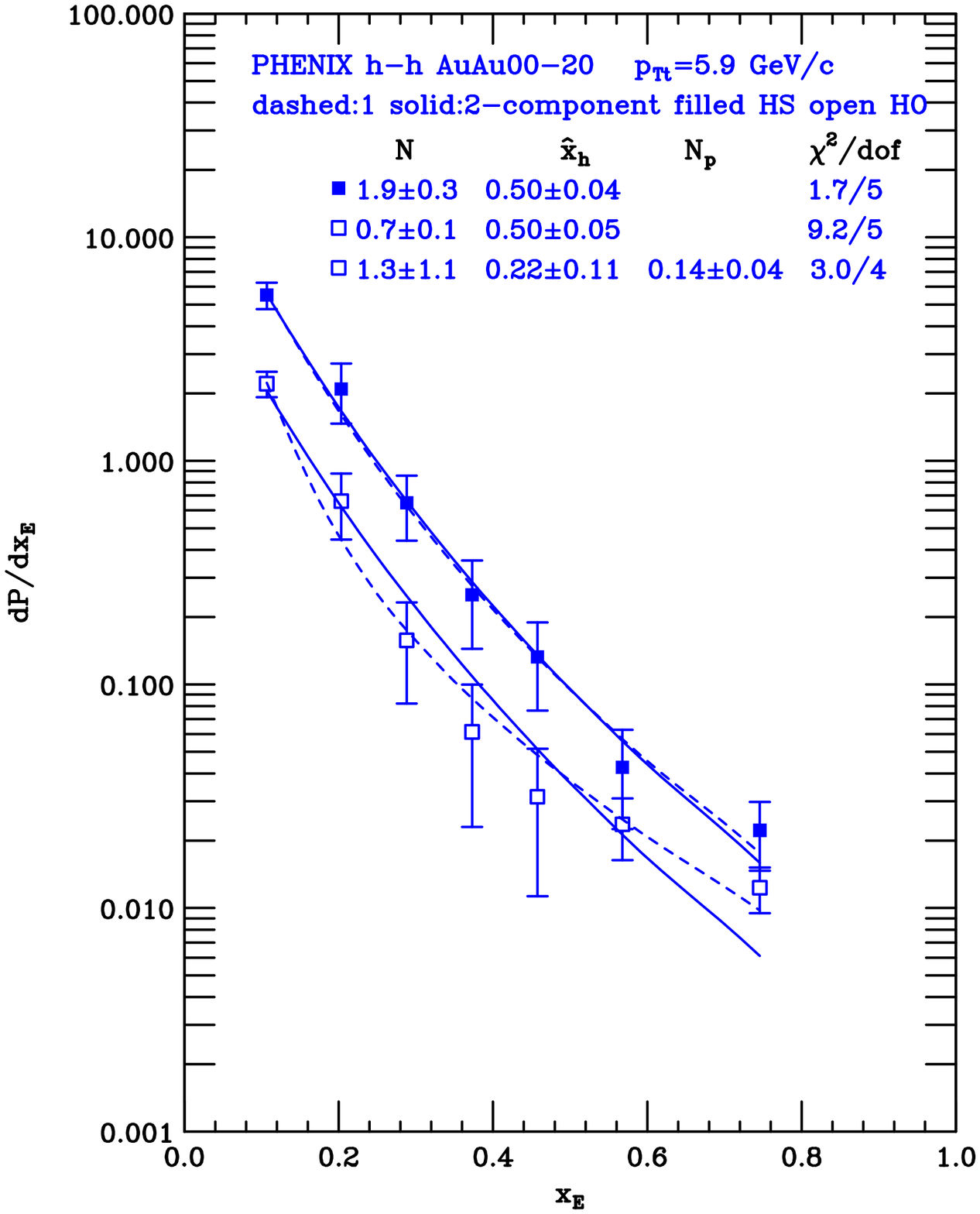}&
\includegraphics[width=0.26\linewidth]{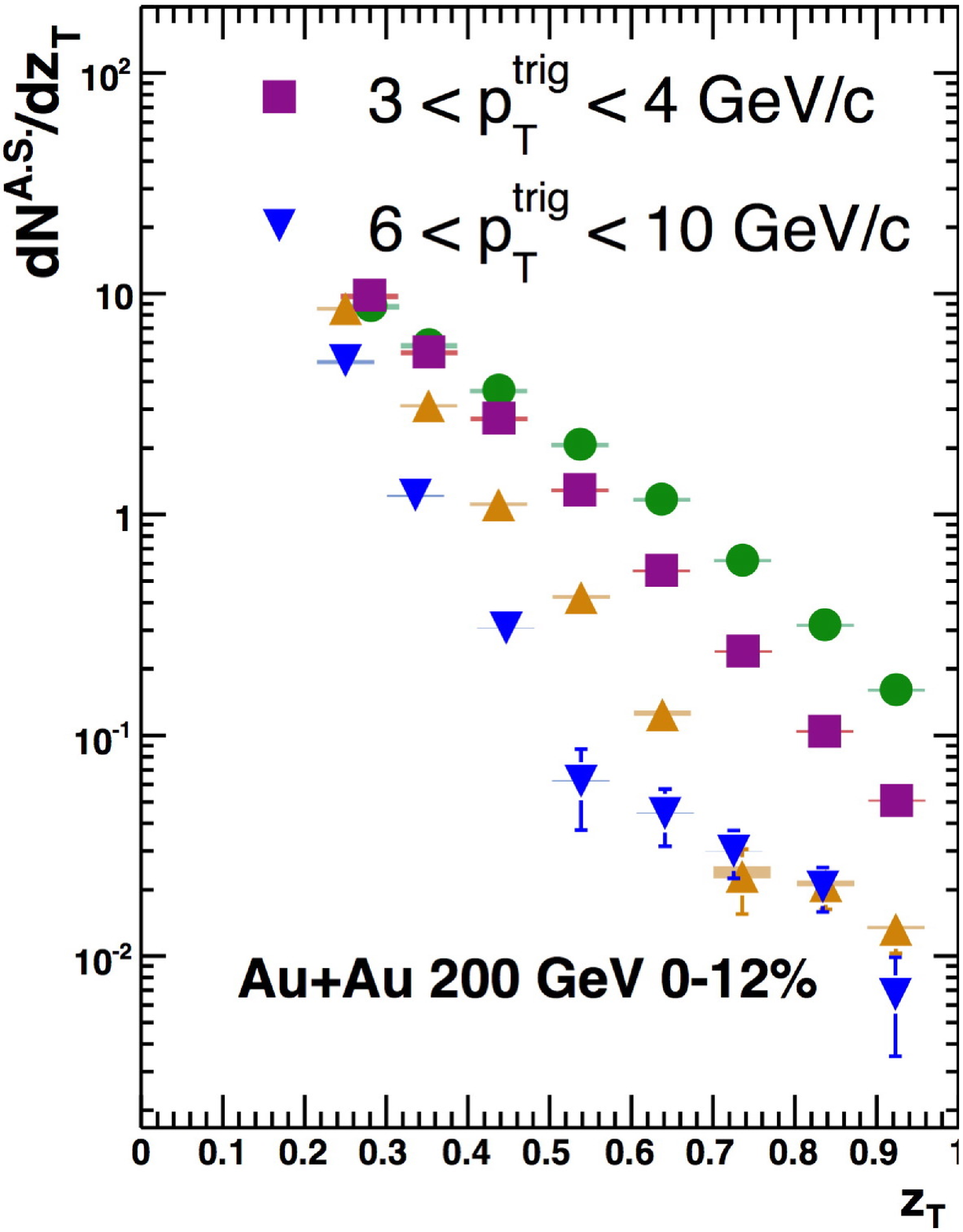}
\end{tabular}
\end{center}\vspace*{-0.12in}
\caption[]{a) (left) Head Region and Shoulder Region definitions~\cite{ppg074}; b) (center) $H+S$, $HO$ data and fit for for AuAu 0-20\%; c) (right) Away-side $x_E=z_T$ distributions for various $p_{T_t}$for AuAu 0-12\%~\cite{Horner}. }
\label{fig:HSdef}\vspace*{-0.12in}
\end{figure}
   
      One of the many interesting new features in Au+Au collisions is that the away side azimuthal jet-like correlations (Fig.~\ref{fig:HSdef}a) are much wider than in p-p collisions and show a two-lobed structure (``the shoulder'') at lower $p_{T_t}$ with a dip at 180$^\circ$,  reverting to the more conventional structure of a peak at 180$^\circ$ (``the head'') for larger $p_{T_t}$. Eq.~\ref{eq:condxe2} provides excellent fits to both these regions in p-p and Au+Au collisions where fits to the Head region (HO) and the full width of the distribution, Head + Shoulder (HS) region, are shown (Fig.~\ref{fig:HSdef}b). The fits give $\hat{x}_h=0.5\pm 0.05$ for Au+Au central collisions for both HO and HS, compared to $\approx 0.8-0.9$ in p-p, clear evidence for energy loss of the away-side parton in the medium produced in Au+Au central collisions. The fit in the Head region (HO) is greatly improved ($\Delta \chi^2=6/1$) if a second component with the same $\hat{x}_h$ as in p-p distributions is added (dashed curve), statistically indicating a parton that has apparently punched through the medium without losing energy, which is evident directly in the STAR measurement~\cite{Horner} (Fig.~\ref{fig:HSdef}c) by the sharp change in slope at $z_T~(x_E)=0.5$  for $6< p_{T_t}< 10$ GeV/c. Clearly, there are lots of new and interesting phenomena to be understood at RHIC. \vspace*{-0.12in}    
\section*{Acknowledgments}
Research supported by U. S. Department of Energy, DE-AC02-98CH10886.

\end{document}